\documentclass[twocolumn, amssymb, amsmath,nobibnotes,nofootinbib, aps, showpacs]{revtex4}
\usepackage[dvips]{graphicx}

\usepackage{graphicx}

\begin{document}

\title{Contrasting electron and hole doping effects on the spin gap of the caged type Kondo semimetal CeOs$_2$Al$_{10}$: A muon spin relaxation and inelastic neutron scattering investigation}

\author{A. Bhattacharyya$^{1,2}$}
\email{amitava.bhattacharyya@stfc.ac.uk} 
\author{D. T. Adroja$^{1,2}$} 
\email{devashibhai.adroja@stfc.ac.uk}
\author{A. M. Strydom$^2$} 
\author{J. Kawabata$^{3}$}
\author{T. Takabatake$^{3}$}
\author{A. D . Hillier$^{1}$}
\author {V. Garcia Sakai$^{1}$}
\author {J. W. Taylor$^{1}$}
\author {R. I. Smith$^{1}$}
\affiliation{$^1$ISIS Facility, Rutherford Appleton Laboratory, Chilton, Didcot Oxon, OX11 0QX, UK} 
\affiliation{$^2$Highly Correlated Matter Research Group, Physics Department, University of Johannesburg, Auckland Park 2006, South Africa}
\affiliation{$^3$Department of Quantum Matter, ADSM and IAMR, Hiroshima University, Higashi-Hiroshima 739-8530, Japan}

\begin{abstract}
The effects of electron (Ir) and hole (Re) doping on the hybridization gap and antiferromagnetic order have been studied by magnetization, muon spin relaxation ($\mu^+$SR), and inelastic neutron scattering on the polycrystalline samples of Ce(Os$_{1-x}$Ir$_x$)$_2$Al$_{10}$ ($x$ = 0.08 and 0.15) and CeOs$_{1.94}$Re$_{0.06}$Al$_{10}$. $\mu^+$SR spectra clearly reveals magnetic ordering below 20 and 10 K for $x$ = 0.08 and 0.15 samples respectively with a very weak signature of oscillations of the muon initial asymmetry at very short time scale. Our important findings are that small amount of electron doping (i) completely suppress the inelastic magnetic excitations near 11 meV down to 2K, which were observed in the undoped compound, and the response transforms into a broad quasielastic response and (ii) the internal field at the corresponding muon site is remarkably enhanced by about ten times compared with the parent compound. On the other hand with small amount of hole (3\% Re) doping the intensity of the inelastic magnetic excitations near 11 meV is reduced significantly. The main origin of the observed doping effect is an extra 5$d$ electrons being carried by Ir and a hole carried by  Re compared with that the Os atom. The obtained results demonstrate a great sensitivity of the carrier doping and provides additional ways to study their anomalous magnetic properties. 
\end{abstract}


\maketitle
\section{Introduction}

Yb or Ce based compounds having elements with $f$-electrons display a rich variety of properties, including an enormous increase of the quasiparticle effective mass (for heavy fermion (HF) system), a Kondo insulating state, and unconventional superconductivity~\cite{Degiorgi,Wachter,Amato}. Over the last two decades these properties attracted considerable interest of condensed matter physics. The qualitative representation of both HF and the Kondo insulating state is based on the knowledge that ground state consequences from a competition between Kondo and Ruderman-Kittel-Kasuya-Yosida (RKKY) interactions~\cite{Doniach}. If the RKKY interaction dominates, the system orders magnetically. However, if the Kondo interaction dominates, theory predicts that hybridization between localized $f$-electron and conduction carrier states should lead to the opening of a charge gap (or pseudogap) at the Fermi energy~\cite{Millis,Georges,Fulde,Hewson,Coleman}. Though this picture is not in dispute, the well-defined predictions of the hybridization scenario have so far escaped direct experimental verification. One of the key predictions is a simple scaling relationship between the magnitude of the direct energy gap $\Delta$ in the excitation spectrum and the enhancement of the effective mass of charge carriers $m^{\star}$ in the coherent regime~\cite{Awasthi}. Several HF materials, such as CeAl$_3$~\cite{Degiorgi1}, show no evidence of such a gap, whereas in prototypical compounds such as UPt$_3$ and URu$_2$Si$_2$ the gap is attributed to a magnetic ground state~\cite{Bonn,Donovan,Sulewski}.

\par
Caged type Ce-based compounds with the general formula CeT$_2$Al$_{10}$ (T = Fe, Ru and Os) have attracted considerable attention due to Kondo semiconducting paramagnetic ground state (down to 40 mK) observed in CeFe$_2$Al$_{10}$~\cite{muro,DTA1} and anomalously high antiferro magnetic (AFM) ordering temperature with spin gap formation at low temperatures in Kondo semimetals CeRu$_{2}$Al$_{10}$ and CeOs$_{2}$Al$_{10}$~\cite{nishi,strydom,muro1}. AFM ordering of these Ce compounds is found to higher than Gd compound which rule out the magnetic order is caused by simple RKKY interaction. According to the de Gennes scaling, $T_N$ (N\'eel temperature) for a Gd compound is expected to be 100 times of that for the Ce counterpart if we neglect the crystal field effect and possible difference in the Fermi surface. Charge$-$density wave like instability is found in optical conductivity measurements for CeT$_2$Al$_{10}$ (T = Ru, Os) which develops along the $b$ axis at temperatures slightly higher than $T_N$. It was suggested that this electronic instability induces the AFM order~\cite{kimura}. The formation of long-range magnetic ordering out of the Kondo semiconducting/semimetallic state itself is unexpected and these are the first examples of this mysterious coexistence of electronic ground states. These compounds also reveal robust anisotropy in magnetic and transport properties, which has been elucidated on the basis of single-ion crystal electric field anisotropy in the presence of strongly anisotropic hybridization between localized 4$f$-electron and conduction electrons. Recently J. Kawabata et.al.~\cite{Kawabata2014} suggested  the suppression of $T_N$ ~\cite{strigari} is well correlated with that of gap energy $\Delta$ as a function of electron/hole doping. They have therefore concluded that the presence of the hybridization gap is in fact necessary for the AFM order at unusually high $T_N$ in CeOs$_2$Al$_{10}$.

\par
In this paper we report the properties of doping of 5$d$ electrons (Ir) and holes (Re) on the Kondo semiconductor CeOs$_2$Al$_{10}$ by measuring magnetization, muon spin relaxation and inelastic neutron scattering on the polycrystalline samples of Ce(Os$_{1-x}$Ir$_x$)$_2$Al$_{10}$ ($x$ = 0.08 and 0.15) and CeOs$_{1.94}$Re$_{0.06}$Al$_{10}$. The broad maxima at around 45 K in magnetic susceptibility for the undoped sample changes to the Curie-Weiss behavior of Ce$^{3+}$ with increasing electron doping. This change means that doping of 5$d$ electrons localizes the 4$f$ electron state in CeOs$_{2}$Al$_{10}$. With hole doping the broad maximum of magnetic susceptibility decreased. The spin-flop transition in $M-H$ for the Ir substituted samples revealed that the direction of the ordered moment changes from $c$-axis to $a$-axis with increasing $x$ to 0.15. In spite of the significant increase of magnetic moment from 0.29 $\mu_B$ for x = 0 to 0.92(3) $\mu_B$ for $x$ = 0.08 per Ce atom~\cite{kato,Khalyavin1}, $T_N$ decreases gradually from 28.5 K to 7.0 K for $x$ = 0 and 0.15 respectively. By the Re substitution, $T_N$ decreases much faster than electron doped samples~\cite{Kawabata2014}.

\begin{table}[htpb]
\begin{center}
\caption{Lattice parameters of Ce(Os$_{1-x}$Ir$_x$)$_2$Al$_{10}$ for $x$ = 0 , 0.08 and CeOs$_{1.94}$Re$_{0.06}$Al$_{10}$ refined from the neutron diffraction data collected at room temperature in the orthorhombic Cmcm space group.}
\begin{tabular}{lccccccccccccccc}
\hline
Compounds && &&  a (\AA) && b (\AA)  && c (\AA) && V (\AA)$^3$ &&\\ 
\hline
CeOs$_2$Al$_{10}$ &&  && 9.1221 && 10.2545 && 9.1694 && 857.729&& \\ 
CeOs$_{1.84}$Ir$_{0.16}$Al$_{10}$ &&  && 9.1193 && 10.2554 && 9.1657 && 857.195 &&\\  
CeOs$_{1.7}$Ir$_{0.3}$Al$_{10}$ &&  && 9.1113 && 10.2544 && 9.1643 && 856.229 &&\\  
CeOs$_{1.94}$Re$_{0.06}$Al$_{10}$ &&  && 9.1217 && 10.2546 && 9.1704 && 857.794  && \\  
\hline
\end{tabular}
\end{center}
\end{table}

\begin{figure}[t]
\vskip 0.4 cm
\centering
\includegraphics[width = 7 cm]{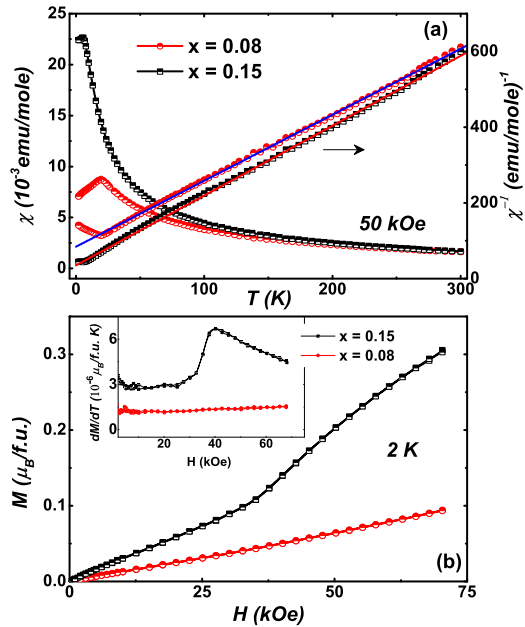}
\caption {(Color online) (a) Temperature dependence of dc suspectibility and inverse $\chi(T)$ of CeOs$_{1.84}$Ir$_{0.16}$Al$_{10}$ and CeOs$_{1.7}$Ir$_{0.3}$Al$_{10}$ in an applied magnetic field 50 kOe. (b) Isothermal field dependence of magnetization at 2 K. Inset shows the first order temperature derivative of $M-H$ data at 2 K.}
\end{figure}

\begin{figure}[t]
\vskip 0.4 cm
\centering
\includegraphics[width = 7 cm]{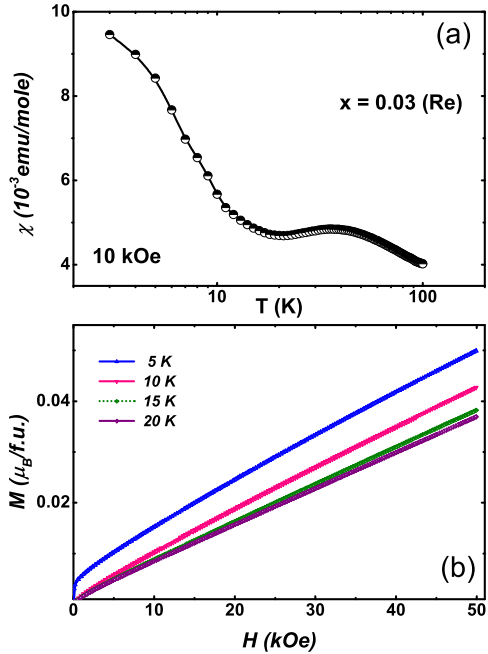}
\caption {(Color online) (a) dc magnetic susceptibility of CeOs$_{1.94}$Re$_{0.06}$Al$_{10}$ in presence of an applied magnetic field 10 kOe. (b) $M~vs.~H$ isotherms at various temperatures.}
\end{figure}

\begin{figure}[t]
\vskip 0.4 cm
\centering
\includegraphics[width = 8 cm]{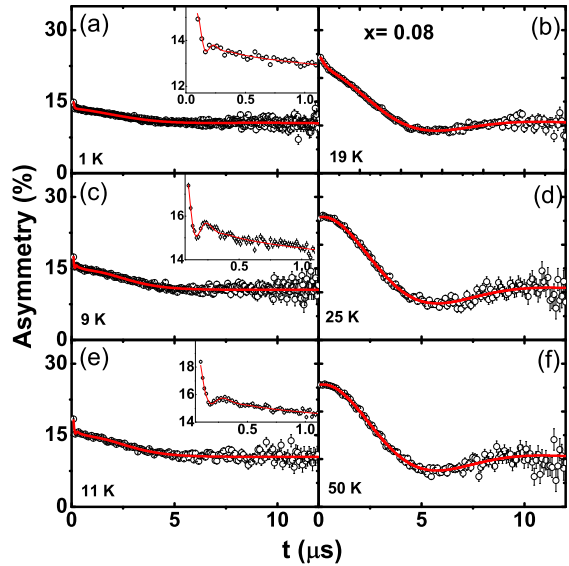}
\caption {(Color online) Zero field time evolution of the muon spin relaxation at various temperatures of CeOs$_{1.84}$Ir$_{0.16}$Al$_{10}$. The solid lines are least-squares fit to the data as described in the text. Insets show the asymmetry spectra in the fast time region.}
\end{figure}

\section{Experimental Details}

Polycrystalline samples of Ce(Os$_{1-x}$Ir$_x$)$_2$Al$_{10}$ ($x$ = 0.08 and 0.15) and CeOs$_{1.94}$Re$_{0.06}$Al$_{10}$ were prepared by arc melting of the constituent elements (Ce : 99.99\%, Os : 99.99\%, Ir : 99.99\%, Re : 99.99\%, Al : 99.999\%) in an argon atmosphere on a water cooled copper hearth. After being  flipped and remelted several times, the boules were wrapped in tantalum foil and annealed at 1123 K for 168 hr under a dynamic vacuum, better than 10$^{-6}$ Torr. Powder x-ray diffraction measurements were carried out using a Panalytical X-Pert Pro diffractometer. Magnetic susceptibility measurements were made using a Quantum Design SQUID magnetometer.
\par
Inelastic neutron scattering (INS) and muon spin relaxation ($\mu^+$SR)  experiments were carried out at the ISIS Pulsed Neutron and Muon Facility of the Rutherford Appleton Laboratory, United Kingdom. To check the phase purity of the samples neutron diffraction measurements were carried out using the GEM time of flight (TOF) diffractometer at ISIS. Powders were loaded into 6mm diameter cylindrical vanadium sample cans and data collected at room temperature for between 30-60 minutes per sample. Inelastic neutron scattering experiments have been performed on the time-of-flight direct geometry chopper spectrometers MARI and MERLIN at the ISIS pulsed neutron facility, UK. The low energy, high resolution, inelastic neutron scattering measurements were also carried out at the ISIS facility using the indirect geometry TOF high resolution spectrometers OSIRIS and IRIS with a pyrolytic graphite (PG) (002) analyser. The INS measurements were carried out between 5 K and 100 K (for CeOs$_{1.7}$Ir$_{0.3}$Al$_{10}$ sample we measured down to 1.2 K on MARI). The powder samples were wrapped in a thin Al foil and mounted inside a thin-walled cylindrical Al can, which was cooled down to 4.5 K inside a top-loading closed cycle refrigerator with He$-$exchange gas around the samples. Incident energies of 6, 25 and 100 meV were used on MARI selected via a Gd-Fermi chopper.  The $\mu^+$SR measurement was carried out on the MUSR spectrometer with the detectors in a longitudinal configuration. The powdered sample was mounted on a high purity silver plate using diluted GE varnish and covered with Kapton film which was cooled down to 1.2 K in a standard He-4 cryostat with He-exchange gas. Spin-polarized muon pulses were implanted into the sample and positrons from the resulting decay were collected in positions either forward or backwards of the initial muon spin direction. The asymmetry is calculated by, $G_z(t) =[ {N_F(t) -\alpha N_B(t)}]/[{N_F(t)+\alpha N_B(t)}]$, where $N_B(t)$ and $N_F(t)$ are the number of counts at the detectors in the forward and backward positions and $\alpha$ is a constant determined from calibration measurements made in the paramagnetic state with a small (20 G) applied transverse magnetic field.

\begin{figure}[t]
\vskip 0.4 cm
\centering
\includegraphics[width = 8 cm]{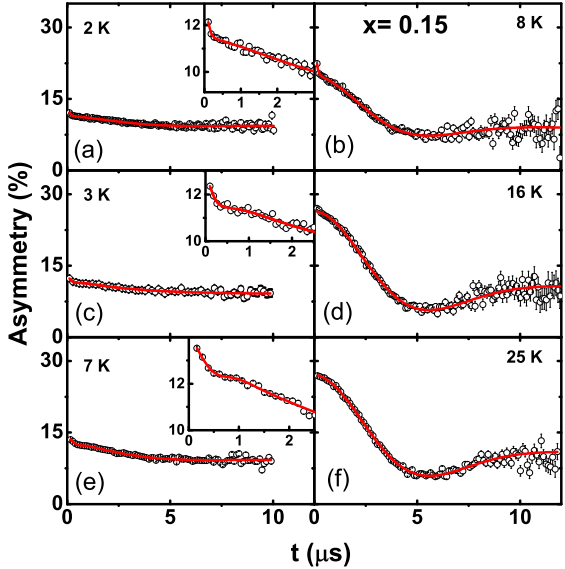}
\caption {(Color online) Time dependence of muon asymmetry spectra at various temperatures of CeOs$_{1.7}$Ir$_{0.3}$Al$_{10}$. The solid line is least-squares fit to the data as described in the text. The insets show the spectra in an early-time region.}
\end{figure}

\section{Structure and Magnetization}

Room temperature neutron diffraction were carried out using GEM diffractometer. Rietveld crystal structure refinement using the neutron diffraction data (not shown here) confirmed that the compounds crystallize in the orthorhombic YbFe$_2$Al$_{10}$-type structure (space group Cmcm, No. 63). In this caged-type structure, the Ce atom is surrounded by a polyhedron formed by 4 Os/Ir and 16 Al atoms and forms a zigzag chain along the orthorhombic $c-$axis.~\cite{Khalyavin1, Khalyavin2014} The refined lattice parameters and  unit cell volume are given in table I. It is clear from neutron diffraction refinement that small amount of electron or hole doping does not change lattice parameters and volume drastically. But for hole doping the ordered moments are substantially reduced (0.18(1)$\mu_B$), while preserving the anomalous magnetic moment direction along the $c$ axis same as CeOs$_2$Al$_{10}$ compound. On the other hand for electron doping moment orient along the $a$ direction, with the ordered value being 0.92(1)$\mu_B$~\cite{Khalyavin1, Khalyavin2014}.

\par
The temperature ($T$) variation of the dc magnetic susceptibility ($\chi = M/H$, where $H$ is the applied magnetic field) measured in zero field cooled condition with $H$ = 50 kOe is shown in Fig. 1 (a).  $\chi$ shows a drop  below 20 K and 6 K for $x$ = 0.08 and 0.15 samples with decreasing temperature. This corresponds to the paramagnetic to antiferromagnetic (PM/AFM) transition in the sample. The magnetic susceptibility of Ce(Os$_{1-x}$Ir$_x$)$_2$Al$_{10}$ ($x$ = 0.08 and 0.15) exhibits Curie-Weiss behavior above 100 K. A linear least-squares fit yields an effective magnetic moment $p_{eff}$= 2.15 and 2.10 $\mu_B$ for the two compositions respectively, which is smaller than free Ce$^{3+}$-ion value (2.54 $\mu_B$), and a negative paramagnetic Curie temperature $\theta_p$= $-$48 and $-$18 K for $x$ = 0.08 and 0.15 samples respectively. The value of magnetic moment suggests that the Ce atoms are in their normal Ce$^{3+}$ valence state. Negative value of $\theta_p$ is indicative of a negative exchange constant. 

\par
Fig. 1 (b) shows the $M$ versus $H$ isotherms recorded at 2 K for $x$ = 0.08 and 0.15.  $M-H$ data imply that the net magnetization in the ordered state of Ce(Os$_{1-x}$Ir$_x$)$_2$Al$_{10}$ is extremely low. It is far from saturation as that expected theoretically ~$gJ$= 2.14 $\mu_B$ for Ce$^{3+}$ ions. This is consistent with previously reported results~\cite{Kawabata2014}. The low values of the observed magnetization is expected for an AFM ground state due to the cancellation of magnetization from different magnetic sublattices of Ce ions. Clear signature of field induced  transition is observed for CeOs$_{1.7}$Ir$_{0.3}$Al$_{10}$ as shown in inset of Fig. 2 (b). 
\par
$\chi(T)$ data of CeOs$_{1.94}$Re$_{0.06}$Al$_{10}$ are shown in Fig. 2 (a) shows PM/AFM transition below $T_N$ = 21 K which matches well with previously reported data~\cite{Khalyavin2014}. Fig. 2 (b) shows the $M$ versus $H$ isotherms recorded at different constant temperatures below and above magnetic transition temperature. $M-H$ data doesn't show any spin flip transition like Ir doped sample. The ordered moments are substantially reduced but preserve the anomalous direction along the $c$ axis. Neutron diffraction results suggest a long-range AFM ordering of the Ce sublattice with a substantially reduced value of the magnetic moment 0.18(1) $\mu_B$ for 3\% Re doped and 0.92(3) $\mu_B$ for 8\% Ir doped, has been found below $T_N$~\cite{Khalyavin2014}.
\begin{figure}[t]
\vskip 0.4 cm
\centering
\includegraphics[width =8 cm]{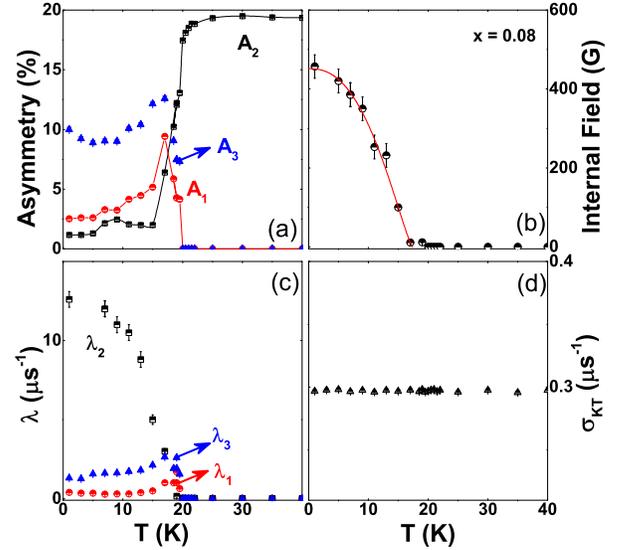}
\caption {(Color online) The temperature dependence of (a) the initial asymmetries $A_1$,  $A_2$ and $A_3$  (b) internal field on the muon site. The solid line in (b) is fit to Eq. 3 (see text). (c) and (d) the depolarization rates $\lambda_1$, $\lambda_2$, $\lambda_3$ and  $\sigma_{KT}$  of CeOs$_{1.84}$Ir$_{0.16}$Al$_{10}$.}
\end{figure}

\begin{figure}[t]
\vskip 0.4 cm
\centering
\includegraphics[width =8 cm]{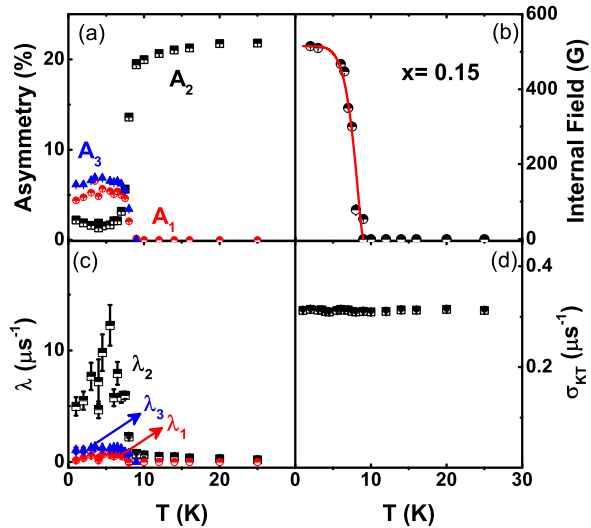}
\caption {(Color online) (a) The initial asymmetries $A_1$,  $A_2$ and $A_3$ (b) internal field on the muon site. The solid line in (b) is fit to Eq. 3 (see text). (c) and (d) the depolarization rates $\lambda_1$, $\lambda_2$, $\lambda_3$ and  $\sigma_{KT}$ of CeOs$_{1.7}$Ir$_{0.3}$Al$_{10}$ as a function of temperature.}
\end{figure}

\section{Muon Spin relaxation}
In order to study the dynamics and estimate distribution of the internal field at the muon site we carried out $\mu^+$SR measurements on Ce(Os$_{1-x}$Ir$_x$)$_2$Al$_{10}$ for $x$= 0.08 and 0.15 samples. Detailed $\mu^+$SR investigations on CeOs$_2$Al$_{10}$ and CeOs$_{1.94}$Re$_{0.06}$Al$_{10}$  were reported previously~\cite{DTA1, Khalyavin2014}. Due to being a local probe, the $\mu^+$SR study provides information on volume fraction of different phases as well as different spin dynamic information for multisite systems.  For small amount of hole doped (3\% Re) compound we found Kubo-Toyabe type behavior above $T_N$ and below $T_N$ its shows clear signature of oscillation (internal field 70 G at base temperature) due to very small magnetic moment (0.18 $\mu_B$)~\cite{Khalyavin2014}.

\par
Figs. 3 (a)-(f) show the zero-field time dependence asymmetry spectra at various temperatures below and above magnetic ordering of CeOs$_{1.84}$Ir$_{0.16}$Al$_{10}$. It is clear from Fig. 3 that small amount of electron doping drastically changes the time dependence of asymmetry spectra compared with the pure compound. Above 20 K, we observe Kubo-Toyabe type behaviour~\cite{Khalyavin2014} i.e. a strong damping at shorter time, and the recovery at longer times, arising from a static distribution of the nuclear dipole moment. In the paramagnetic state i.e. above 20 K we used the following function to fit our $\mu^+$SR spectra as shown in Figs. 3 (d) and (f): 
\begin{equation}
G_{z_1}^{KT}(t) =A_2 e^{-\lambda_2 t}KT(t)+A_{bg}
\end{equation}
where $KT(t) = \frac{1}{3}[1+2(1-\sigma_{KT}^2t^2)e^{(-\frac{\sigma_{KT}^2t^2}{2})}$] is the well-known  static Kubo-Toyabe (KT) function.  Initial amplitude of the KT decay is $A_2$;  $\lambda_2$ is the electronic relaxation rate; $A_{bg}$ is a constant background arising from muon stopping on the silver sample holder. $A_{bg}$ was estimated from 50 K data and kept fixed for fitting all the other spectra. Nuclear depolarization rate is $\sigma_{KT}$, $\sigma_{KT}/\gamma_{\mu}$ = $\Delta$ is the local Gaussian field distribution width, $\gamma_{\mu}$ is the gyromagnetic ratio of the muon. $\sigma_{KT}$ was found to be almost temperature independent as shown in Fig. 5 (d) with its value equal to 0.29 $\mu s^{-1}$. Using a similar $\sigma_{KT}$ value  Kambe {\it et. al.}~\cite{S. Kambe} have suggested $4a$ as the muon stopping site in CeRu$_2$Al$_{10}$, while for CeOs$_2$Al$_{10}$~\cite{DTA1}, the muon stopping site was assigned to the (0.5, 0, 0.25) position. Very recent, Guo {\it et. al.}~\cite{HG} have investigated Ce(Ru$_{1-x}$Rh$_x$)$_2$Al$_{10}$ using $\mu^+$SR and their dipolar fields calculation supports the (0.5, 0, 0.25) site for the muons.

Our neutron diffraction result~\cite{Khalyavin1} shows the value of ordered magnetic moment is 0.92 $\mu_B$/Ce for  CeOs$_{1.84}$Ir$_{0.16}$Al$_{10}$ which is  substantially bigger compared to parent compound [0.29($\pm$0.05) $\mu_B$/Ce observed in the parent compound CeOs$_{2}$Al$_{10}$~\cite{kato}]. In contrast to our undoped compound which shows three frequencies below magnetic ordering here we observe very weak signature of oscillation. Muon spin precession is weak due to the fact that internal field exceeds the maximum internal field ($\approx$ 800 G) detectable on the MUSR spectrometer due to the pulse width of the ISIS muon beam. This supports the findings of our neutron diffraction data~\cite{Khalyavin1}. As the frequency response goes to very low time region the fitting procedure is not straightforward, and it becomes difficult to determine the fitting parameters unambiguously.
\par
Below 20 K, all spectra are described uniformly by the phenomenological function
\begin{equation}
G_{z_2}(t) = A_1cos(\omega t+\phi)e^{-\lambda_1 t}+ A_3 e^{-\lambda_3 t} + G_{z_1}^{KT}(t)
\end{equation}

where $\lambda_i$ ($i$ = 2 and 3) is the muon depolarization rate (arising from the distribution of the internal field), $\phi$ is the phase and $\omega$ = $\gamma_\mu H_{int}$ is the muon precession frequency ($H_{int}$ is the internal field at the muon site). The first term represents the transverse components of the internal fields seen by the muons along which they precess, while the second term represents the longitudinal component. The last term in Eq. (2) is the Kubo-Toyabe term, which accounts for the dip seen in the $\mu^+$SR spectra near 5.5 $\mu s$ even in the magnetically ordered state, indicating that the internal field seen by the muons on this site is smaller than that of the nuclear field. This suggests that the corresponding muon site can be assigned to the $4a$ crystallographic position~\cite{HG}.

\begin{figure}[t]
\vskip 0.4 cm
\centering
\includegraphics[width = 6 cm]{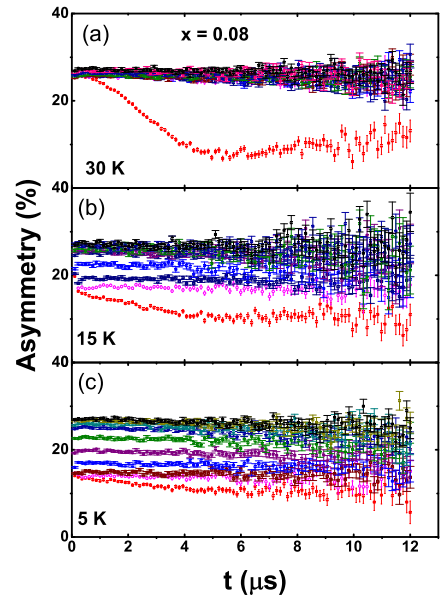}
\caption {(Color online) (a), (b) and (c) represent the $\mu^+$SR spectra for various applied magnetic fields ($H$ = 0, 50, 100, 250, 500, 1000, 2000, 3000, 4000	and 4500 G) at 30 K, 15 K and 5 K respectively for CeOs$_{1.84}$Ir$_{0.16}$Al$_{10}$.}
\end{figure}

\begin{figure}[t]
\vskip 0.4 cm
\centering
\includegraphics[width = 8 cm]{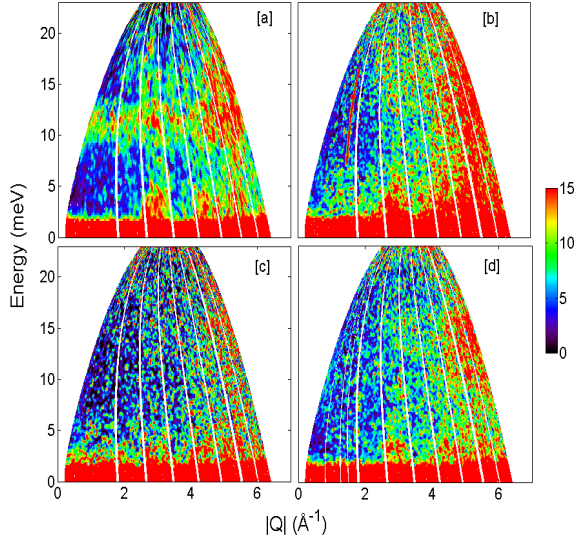}
\caption {(Color online) Inelastic neutron scattering intensity of (a) CeOs$_{2}$Al$_{10}$ (b) CeOs$_{1.84}$Ir$_{0.16}$Al$_{10}$, (c) CeOs$_{1.7}$Ir$_{0.3}$Al$_{10}$ and (d) CeOs$_{1.94}$Re$_{0.06}$Al$_{10}$ at 5 K, measured with respective incident energy of E$_i$= 25 meV on the MARI spectrometer.}
\end{figure}

\begin{figure}[t]
\vskip 0.4 cm
\centering
\includegraphics[width = 7 cm]{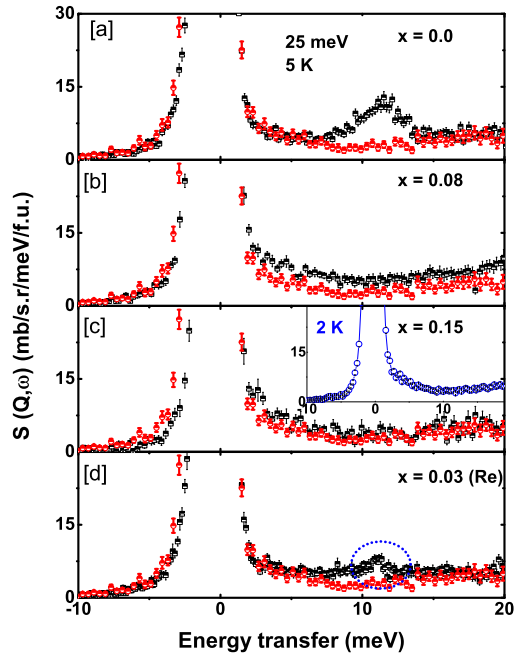}
\caption {(Color online)  $Q$ integrated (0$\le$$Q$$\le$2.5~\AA) intensity versus energy transfer of (a) CeOs$_{2}$Al$_{10}$ (b) CeOs$_{1.84}$Ir$_{0.16}$Al$_{10}$, (c) CeOs$_{1.7}$Ir$_{0.3}$Al$_{10}$, (d) CeOs$_{1.94}$Re$_{0.06}$Al$_{10}$ along with LaOs$_2$Al$_{10}$ (red circle)  at 5 K, measured with respective incident energie of E$_i$= 25 meV. Inset shows INS spectra at 2 K for CeOs$_{1.7}$Ir$_{0.3}$Al$_{10}$.}
\end{figure}

\begin{figure}[t]
\vskip 0.4 cm
\centering
\includegraphics[width = 7.5 cm]{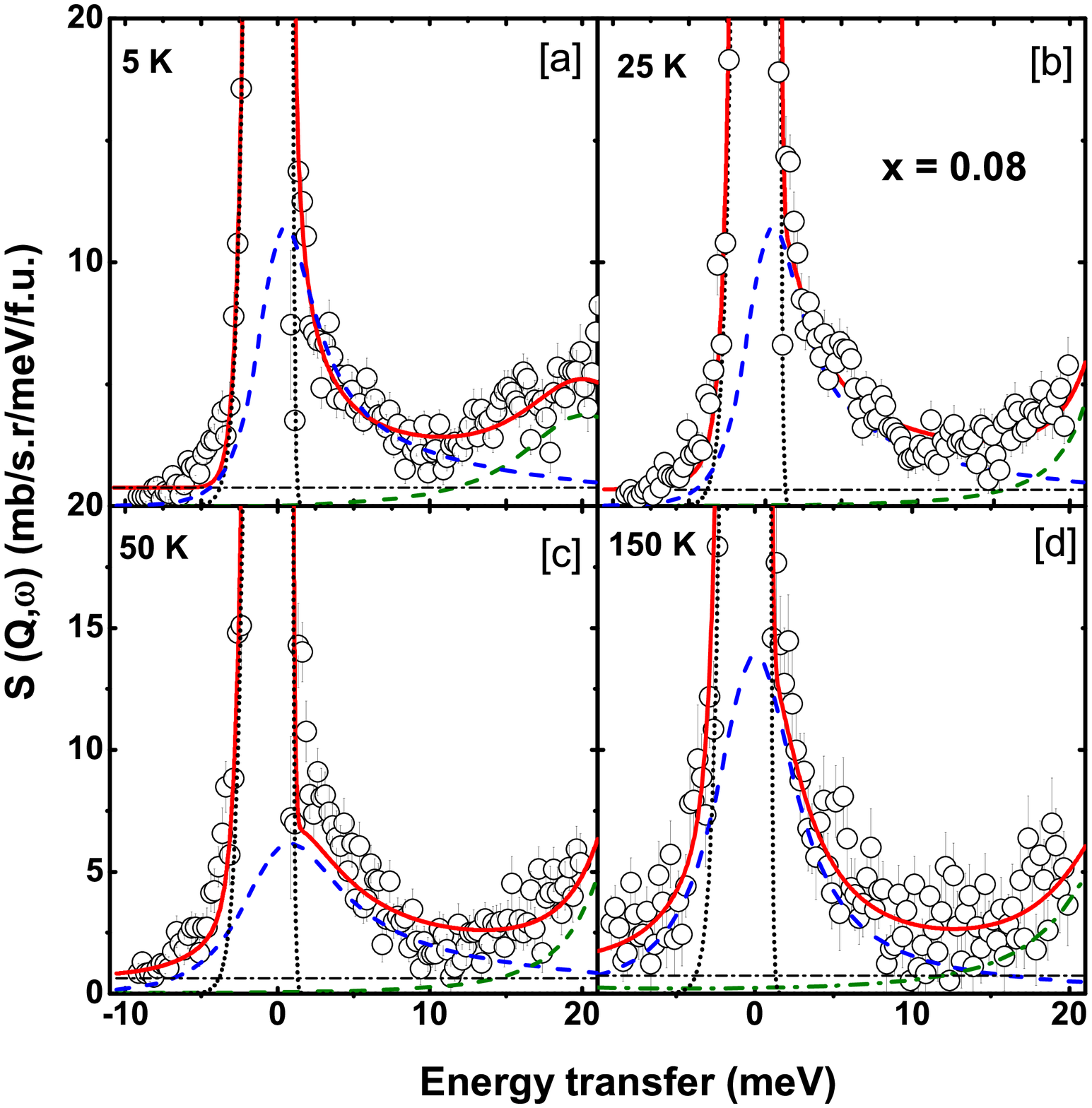}
\caption {(Color online) $Q$ integrated (0$\le$$Q$$\le$2.5~\AA) intensity versus energy transfer of CeOs$_{1.84}$Ir$_{0.16}$Al$_{10}$  at 5, 25, 50 and 150 K respectively measured with respective incident energy of E$_i$= 25 meV. The solid line represents the fit using an inelastic peak (dash-dotted line represents the components of fit), and the dotted line represents a quasielastic peak.}
\end{figure}

\begin{figure}[t]
\vskip 0.4 cm
\centering
\includegraphics[width = 7.5 cm]{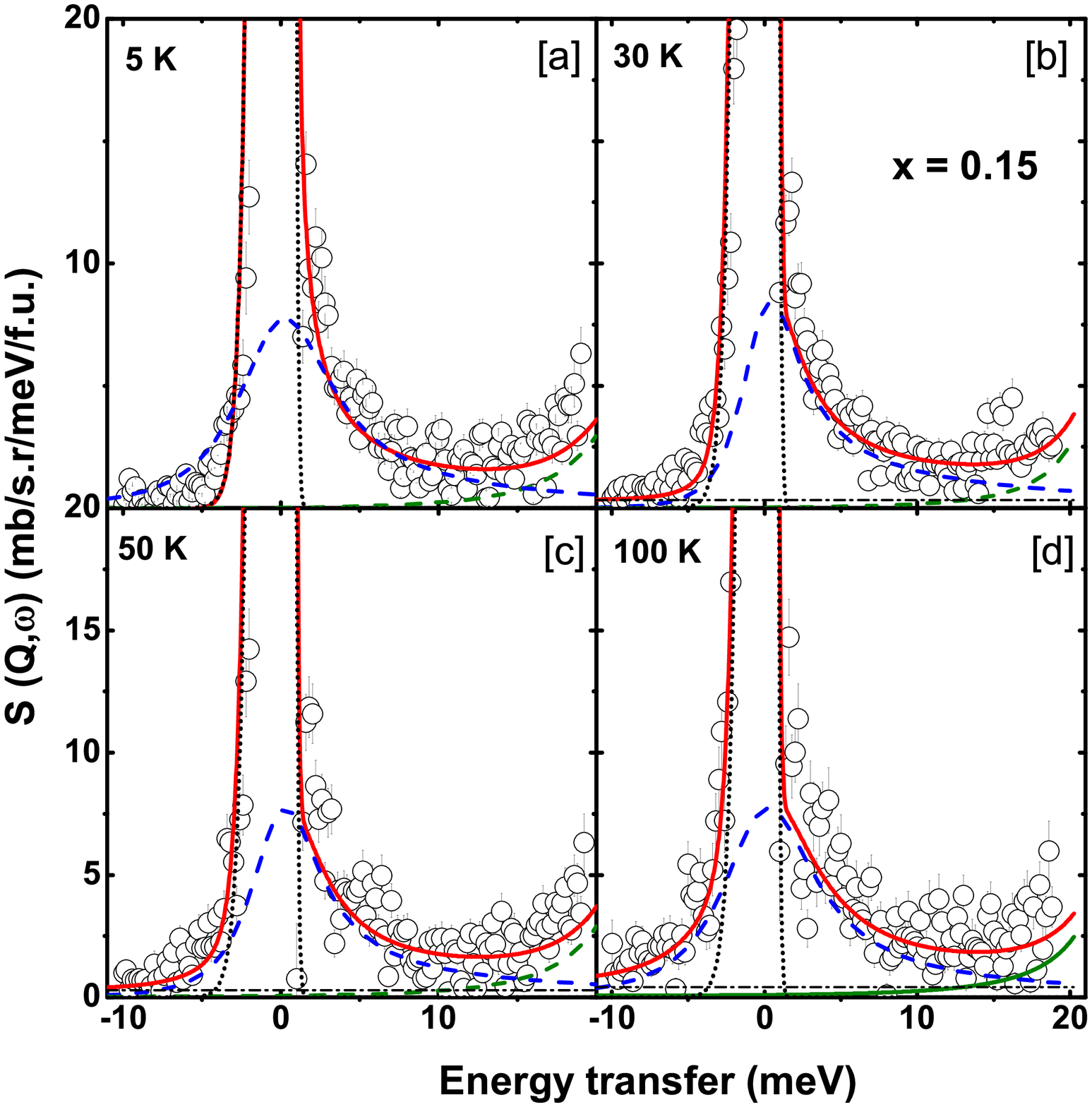}
\caption {(Color online) $Q$ integrated (0$\le$$Q$$\le$2.5~\AA) intensity versus energy transfer of CeOs$_{1.7}$Ir$_{0.3}$Al$_{10}$  at 5, 30, 50 and 150 K respectively measured with respective incident energy of E$_i$= 25 meV. The solid line represents the fit using an inelastic peak (dash-dotted line represents the components of fit), and the dotted line represents a quasielastic peak.}
\end{figure}

\begin{figure}[t]
\vskip 0.4 cm
\centering
\includegraphics[width = 6.5 cm]{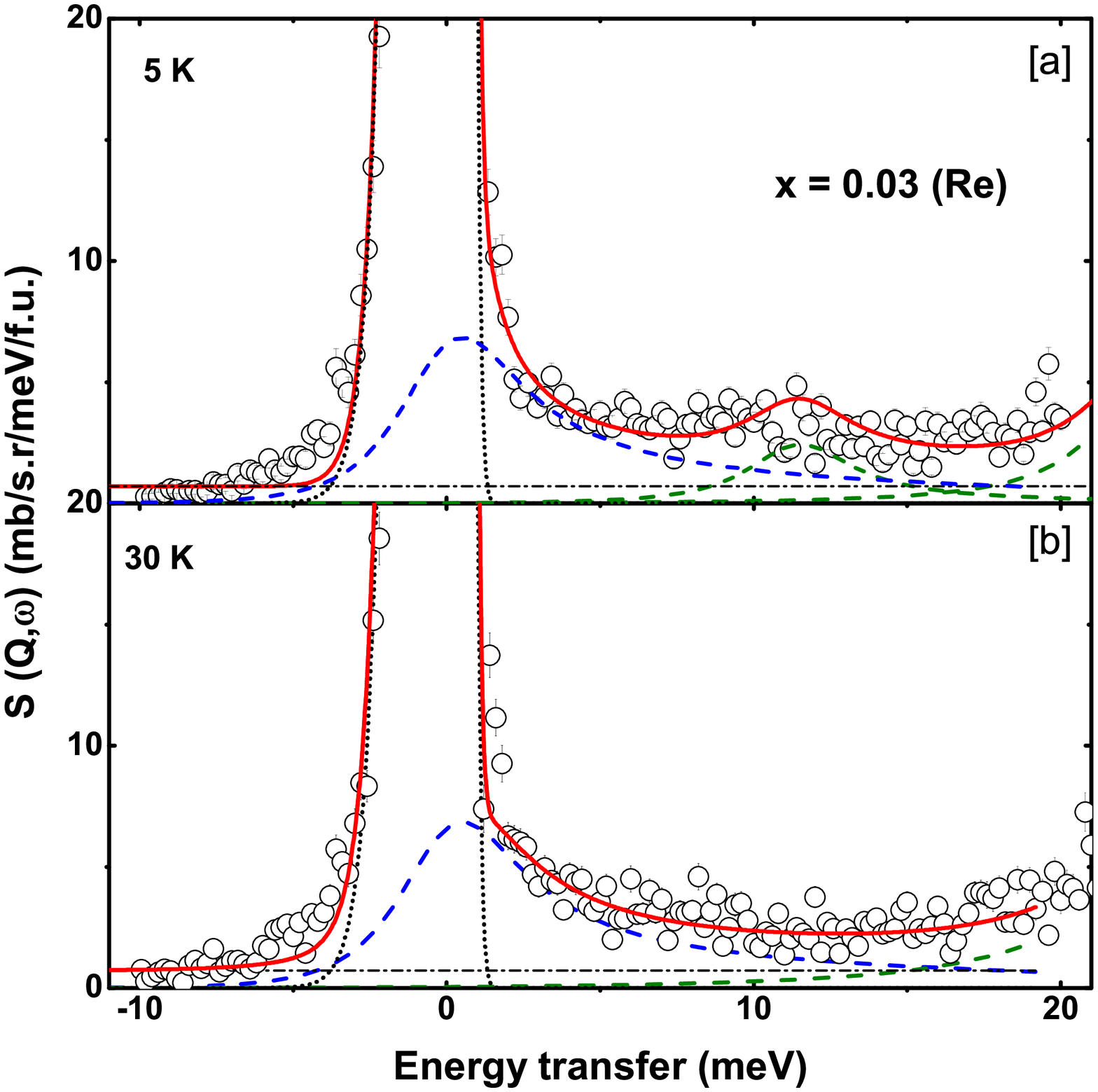}
\caption {(Color online) $Q$ integrated (0$\le$$Q$$\le$2.5~\AA) intensity versus energy transfer of CeOs$_{1.94}$Re$_{0.06}$Al$_{10}$  at 5 and 30 K measured with respective incident energy of E$_i$= 25 meV. The solid line represents the fit using an inelastic peak (dash-dotted line represents the components of fit), and the dotted line represents a quasielastic peak.}
\end{figure}

\begin{figure}[t]
\vskip 0.4 cm
\centering
\includegraphics[width = 6 cm]{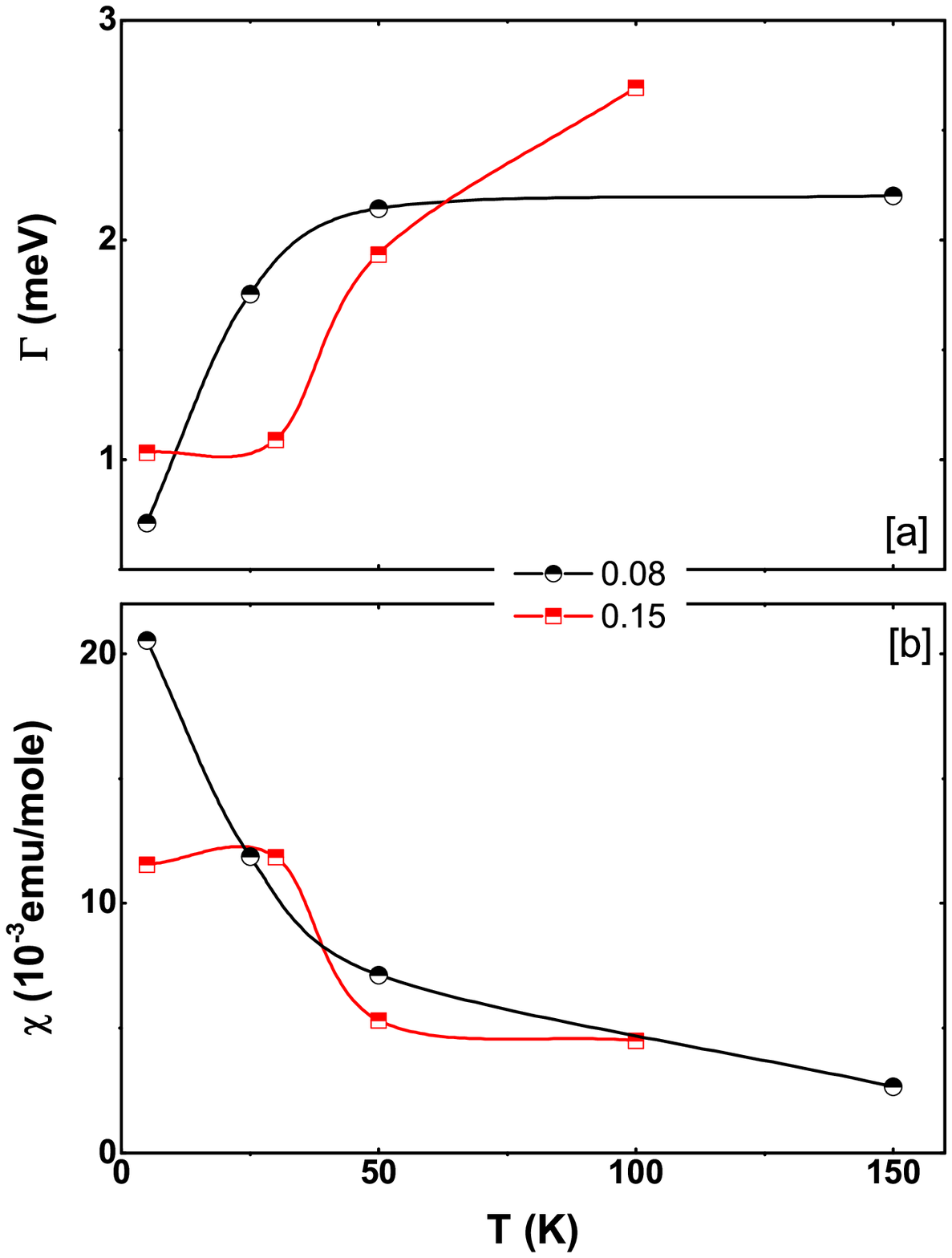}
\caption {(Color online) (a) and (b) represent the  variation of linewidth and susceptibility as a function of temperature obtained from fitting
the magnetic scattering intensity of Ce(Os$_{1-x}$Ir$_{x}$)$_2$Al$_{10}$ ($x$ = 0.08 and 0.15). Lines are guides to the eye.}
\end{figure}

\par
The temperature dependencies of these parameters are shown in Figs. 5 (a)-(d). Below 20 K, as shown in Fig. 5 (a) there is a loss of initial asymmetry in $A_2$ that of the high temperature value. The initial asymmetry associated with frequency term $A_1$ starts to increase below this temperature (20 K) [see Fig. 5 (a)], indicating the onset of a long-range ordered state in CeOs$_{1.84}$Ir$_{0.16}$Al$_{10}$ which agrees with the specific heat and magnetic susceptibility data~\cite{Kawabata2014}. The temperature dependence of Lorentzian decay term $\lambda_2$ starts to increase below 20 K. $\lambda_1$ associated with oscillating term almost remains constant. Fig. 5 (d) shows the temperature dependence of the muon depolarization rate which seems temperature independent. Fig. 5 (b) shows the temperature dependence of the internal field (or muon precession frequency) at the muon site. This shows that the internal fields appear below 20 K, signifying clear evidence for long-range magnetic order. However, the associated internal fields (approximately 500 G) are found to be one order of magnitude larger than in compared to CeOs$_2$Al$_{10}$ ($H_{int}$ = 40 G at base temperature) compound, which indicates a large ordered state magnetic moment of the Ce$^{3+}$ ion. This observation is in accordance with our recent neutron diffraction results~\cite{Khalyavin1}. 

\par
Figs. 4 (a)-(f) show the zero field asymmetry spectra for CeOs$_{1.7}$Ir$_{0.3}$Al$_{10}$ sample. The Kubo-Toyabe behavior is observed above 10 K, while a loss of initial asymmetry is observed below 10 K. However, only a fast precession frequency is observed. The spectra exhibit similar behaviors which suggests the same ground state for $x$ = 0.08 and 0.15 samples. Temperature dependence of the extracted fitting parameters (using Eqns. 1 and 2) for $x$ = 0.15 are shown in Figs. 6 (a)- (d). 

\par

In order to find out the nature of the magnetic interaction for $x$ = 0.08 and 0.15 (Ir doped) temperature dependence of internal field was fitted
\begin{equation}
H_{int}(T)= H_0\left(1-\left(\frac{T}{T_N}\right)^{\alpha}\right)^{\beta}
\end{equation}

Observed parameters are $\beta$ = 0.45(2) and 0.47(2), $H_0$ = 451 G and 510 G, $\alpha$ = 2.46 and 4.4 and $T_{N}$ = 18.5(3) K and 8.2(2) K (see Figs. 5 (b) and 6 (b)) for $x$ = 0.08 and 0.15 respectively. A good fit with $\beta \sim$ 0.45 (or 0.47) suggests the magnetic interactions in electron doped system is long-range spin-spin interactions. $\alpha >$  1 indicates complex magnetic interactions in this system.~\cite{dikh,sb} 

\par
To estimate the internal field at the muon site at low temperature we did field dependence measurements. Figs. 7 (a)-(c) show the effect of applied magnetic field on the $\mu^+$SR spectra of CeOs$_{1.84}$Ir$_{0.16}$Al$_{10}$ at (i) 5 K and 15 K in the magnetic ordered state (ii) 30 K in the paramagnetic state. Zero field $\mu^+$SR spectra  at 5 K, reveal a strong suppression of the KT term due to a broad distribution of internal fields.  On increasing the field to 100 G, the KT term disappears but the full instrumental asymmetry is not recovered until an applied field of 2500 G. This may suggest that the internal fields at the muon sites are below 2500 G. For CeOs$_2$Al$_{10}$ we observed an internal field less than 100 G from field dependence $\mu^+$SR spectra. The Kubo-Toyabe term is decoupled in a field 50 G at 30 K and the spectra are nearly time independent as shown in Fig. 7 (a).

\section{Inelastic Neutron scattering}

INS measurements on CeOs$_{2}$Al$_{10}$ clearly reveal the presence of a sharp inelastic magnetic excitation near 11 meV (as shown in Figs. 8(a) and 9 (a)) between 5 and 26 K, due to opening of a gap in the spin-excitation spectrum, which transforms into a broad response at and above 30 K ~\cite{DTA1}. Small amount of electron and hole doping results dramatic change in zero field $\mu^+$SR spectra. It is great interest to study inelastic neutron scattering to see how the spin gap and its $Q$ and temperature dependence varies with doping. In this section we briefly report the temperature dependent of low energy INS spectra of Ce(Os$_{1-x}$Ir$_x$)$_2$Al$_{10}$ ($x$ = 0.08 and 0.15) and CeOs$_{1.94}$Re$_{0.06}$Al$_{10}$. We have also measured the nonmagnetic phonon reference compound LaOs$_2$Al$_{10}$. INS investigations on CeT$_2$Al$_{10}$ (T=Fe, Ru, and Os) compounds were previously reported by several groups~\cite{DTA1,DTA2,jmm,jmm1}.

\par

Figs. 8 (a) - (d) show the plots of the total scattering intensity (magnetic and phonon contributions), energy transfer {\it vs.} momentum transfer measured at 4.5 K on the MARI spectrometer for $x$ = 0, 0.08, 0.15 and CeOs$_{1.94}$Re$_{0.06}$Al$_{10}$. For accurate estimation of magnetic scattering in the Ce compounds, we subtracted the phonon reference by using the data of nonmagnetic reference compound LaOs$_2$Al$_{10}$. We scaled the La data by the cross-section ratio of the Ce compounds and LaOs$_2$Al$_{10}$ and then subtracting from the Ce data (we call this as method 1)~\cite{dta3}. For undoped compound ($x$ = 0) we find clear magnetic excitation or spin gap energy around 11 meV~\cite{DTA1}. This value is in agreement with the spin gap energy estimated from  the exponential behavior of the observed magnetic susceptibility, specific heat, and NMR studies~\cite{csl}. In the AFM ordered state for $x$ = 0.08 and 0.15 samples the excitation disappears and the response transforms into a broad quasielastic or inelastic feature as shown in Figs. 9 (b) - (c). On the other hand for CeOs$_{1.94}$Re$_{0.06}$Al$_{10}$ we find weak signature of magnetic excitation around 11 meV but the intensity is quite reduced compared to the pure compound. Further for comparison we have also shown the scattering from the nonmagnetic reference compound LaOs$_2$Al$_{10}$ which reveals the magnetic natures of the excitations in the Ce compounds. Since $x$ = 0.15 sample orders below 6 K, we measured down to 2 K to check any signature of a spin gap. But it is clear from inset of Fig. 9 (c) that there is no clear sign of spin gap. 

\par

Now we discuss the temperature dependence of the spin-gap excitation energy. Magnetic scattering at various temperatures for  CeOs$_{1.84}$Ir$_{0.16}$Al$_{10}$, CeOs$_{1.7}$Ir$_{0.3}$Al$_{10}$ and CeOs$_{1.94}$Re$_{0.06}$Al$_{10}$ are shown in Figs. 10 -13.  We used the ratio of the high-$Q$ and low-$Q$ data of LaOs$_2$Al$_{10}$ (i.e., ratio = $[S_{high}(Q,\omega)/S_{low}(Q,\omega)]_{La}$) to estimate the magnetic scattering: $S_M(Q,\omega) = S_{low}(Q, \omega)_{Ce}-S_{high}(Q, \omega)_{Ce}/ratio$ (we call this as method 2)~\cite{dta3}. We have plotted the data in one-dimensional (1D) ($Q$-integrated between 0 and 2.5 \AA) energy cuts (see Figs. 10-12) taken from the two-dimensional (2D) plots to see the linewidth ($\Gamma$) and intensity clearly. In order to investigate the involvement of prevailing inelastic-type  energy excitations, we analyzed the temperature dependent magnetic scattering, $S({Q,\omega})$ using a Lorentzian lineshape~\cite{DTA1}, and fits are shown in Figs. 10$-$12 for all three samples. Fig. 13 shows the temperature-dependent parameters estimated from the fit to the data for $x$ = 0.08 and 0.15. (solid lines are INS fits). Temperature dependent of linewidth, $\Gamma$, for $x$ =  0.08 and 0.15 are shown in Fig. 13 (a). The quasielastic linewidth is a measure of Kondo temperature $T_K$ (just above $T_N$, ideally one takes the $\Gamma$ value at $T$= 0). $T_K$ value is found to be 22 K and 10 K for $x$ = 0.08 and 0.15 compounds respectively. The smaller value of $T_K$ indicates that Ce ion are nearly localized in the doped systems compared to undoped parent compound. Fig. 13 (b) shows the estimated magnetic susceptibility for both compounds which matches well with measured $\chi(T)$ data. Thus the van Vleck contribution from the high-energy crystal electric field (CEF) is small at low-$T$.

\section{SUMMARY}

A comparative $\mu^+$SR and inelastic neutron scattering study has been performed on the caged type compounds Ce(Os$_{1-x}$Ir$_x$)$_2$Al$_{10}$ ($x$ = 0.08 and 0.15) and CeOs$_{1.94}$Re$_{0.06}$Al$_{10}$ to understand the unusual AFM phase transition and spin-gap formation. Our $\mu^+$SR spectra for $x$ = 0.08 and 0.15 clearly reveal the presence of the long-range magnetic ordering at low temperature. One muon frequency is observed for doped samples in contrast to three frequencies of parent CeOs$_2$Al$_{10}$ compound. The internal field at the corresponding muon site is enhanced by about ten times by small amount of electron doping, which supports the larger moment being along the $a$ axis in the Ir-doped sample as revealed by our recent neutron diffraction experiment. The AFM structure of CeOs$_2$Al$_{10}$ is not so robust and can be easily tuned by external perturbations such as chemical doping. The important results is that small amount of electron doping completely suppress the inelastic excitation near 11 meV, which is present in the undoped compound.. The main origin of the observed doping effect is the extra 5$d$ electrons carried by Ir. Our study has revealed important new results in the wider framework of spin-gap formation driven by 4$f$ and conduction electrons’ hybridization. Further study of single crystals of Ce(Os$_{1-x}$Ir$_x$)$_2$Al$_{10}$ are highly desirable using low as well as high energy inelastic neutron scattering, also in polarized mode, to confirm the disappearance of spin gap in the electron doped systems.

\section*{ACKNOWLEDGEMENT}
We would like to thank P. Deen, D. D. Khalyavin and W. Kockelmann for interesting discussions. D.T.A., A.B thanks the FRC of UJ and ISIS-STFC for funding support. D.T.A., and A.D.H. would like to thank CMPC-STFC, grant number CMPC-09108, for financial support.  A.M.S. thanks the SA-NRF (Grant 78832) and UJ Research Committee for financial support. T.T. thanks KAKENHI No.26400363 from MEXT, Japan.

\end{document}